\documentstyle[twoside,fleqn,espcrc2]{article}

\newcommand{\oo}{\"{o}}
\newcommand{\neu}{\tilde{\chi}}
\def\lsim{\mathrel{\rlap{\lower4pt\hbox{\hskip1pt$\sim$}}
    \raise1pt\hbox{$<$}}}         
\def\gsim{\mathrel{\rlap{\lower4pt\hbox{\hskip1pt$\sim$}}
    \raise1pt\hbox{$>$}}}         

\begin{document}

\title{Neutrino-induced Muon Fluxes from Neutralino Annihilations in the
Sun and in the Earth}

\author{J.~Edsj{\oo}\address{Department of Theoretical Physics, Uppsala
University, Box 803, S-751 08 Uppsala, Sweden}\thanks{E-mail:
edsjo@teorfys.uu.se}}

\begin{abstract}
The flux of neutrino-induced muons at the surface of the
Earth is calculated from injection of neutralino annihilation products in the
core of the Sun and the  Earth. An improved treatment of neutrino propagation
through the Sun is performed and the results are presented in an easy-to-use
parameterization. For an explicit supersymmetric model, an observable
neutralino annihilation signal is demonstrated.
\end{abstract}

\maketitle


\section{Introduction}

The dark matter in the Universe may be constituted by the neutralino, which
probably is the lightest supersymmetric particle. If these particles exist
they
will get
trapped gravitationally by the Earth and the Sun where they can annihilate
and produce other particles, e.g.\ fermion--antifermion pairs, gauge bosons and
Higgs bosons. These particles can hadronise and/or decay producing high
energy muon neutrinos which can be detected in neutrino detectors, in which
the neutrinos produce muons via charged current interactions and
the muons can be detected due to the \v{C}erenkov light they emit.

We have considered the whole chain of processes from the annihilation
products in the core of the Sun or the Earth to detectable muons at the
surface of the Earth, see \cite{Edsjo} for details. The neutrino
flux at the surface of the Earth has been calculated earlier by others
, e.g.\ \cite{RS,Kamionkowski,Kamiokande}, but in this calculation the neutrino
propagation through the Sun has been considered more carefully and the neutrino
interactions near the detector have been simulated to get the muon fluxes.

The hadronisation and/or decay of the  annihilation products have been
simulated with {\sc Jetset} 7.3 \cite{Jetset} and the neutrino interactions
on the way out of the Sun have been simulated with {\sc Pythia} 5.6
\cite{Jetset}. The neutrino interactions near the detector (in which the
detectable muons are produced) have also been simulated with {\sc Pythia} 5.6.


\begin{figure*}[htb]
  \vspace{4.4cm}
  \includegraphics{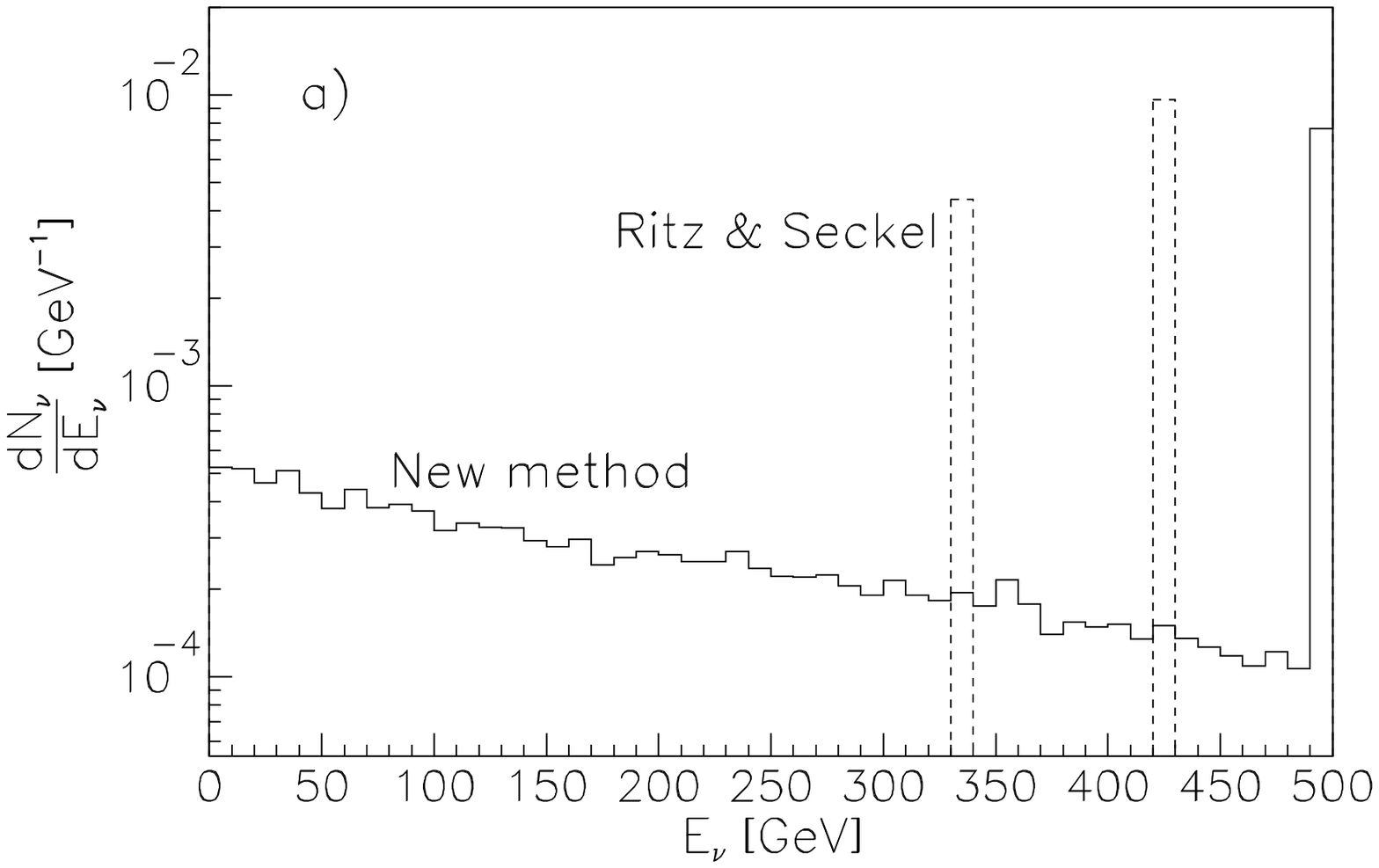}
  \includegraphics{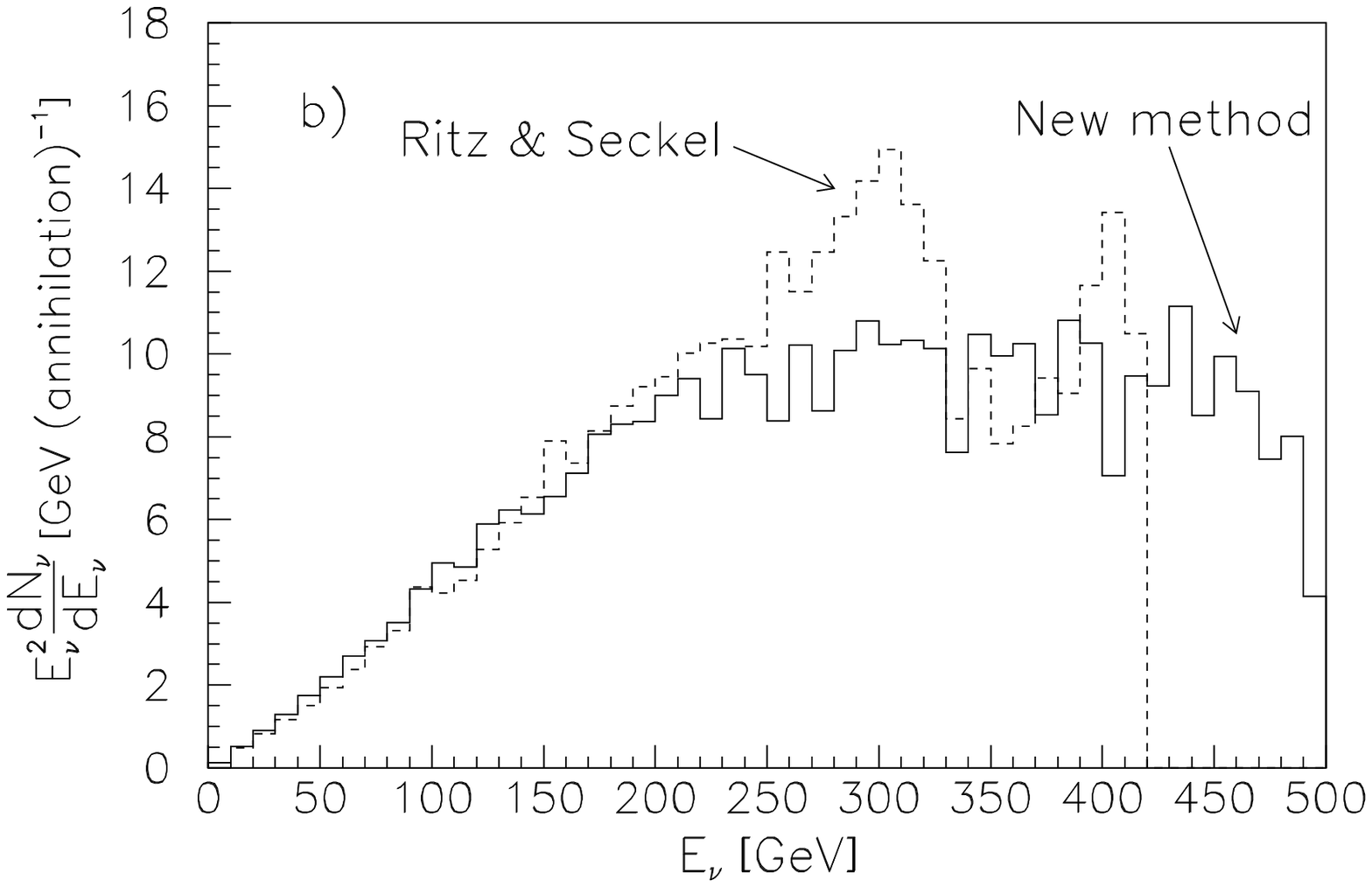}

  \caption{a) The neutrino spectrum ($\nu_{\mu} + \bar{\nu}_{\mu}$) at the
  surface of the Sun from
  injection of 500 GeV neutrinos and anti-neutrinos in the core of the
  Sun with the method of Ritz and Seckel \protect\cite{RS} and with the new
  method. b) The neutrino spectrum ($\nu_{\mu} + \bar{\nu}_{\mu}$) weighted by
  $E_{\nu}^2$ at the surface of
  the Sun for the $W^+ W^-$-channel with a neutralino mass of 500
  GeV\@. The two peaks in the Ritz and Seckel spectra correspond to $\nu_{\mu}$
  and $\bar{\nu}_{\mu}$ respectivley.}
  \label{fig:enucomp}
\end{figure*}

\section{Annihilation channels and interactions of annihilation
products}

A pair of neutralinos can annihilate to produce a fermion-antifermion
pair or gauge bosons, Higgs bosons and gluons, i.e.
$\neu \neu \to \ell^+\ell^-$, $q\bar{q}$, $gg$, $q\bar{q}g$, $W^+W^-$,
$Z^0Z^0$, $Z^0H^0$, $W^{\pm}H^{\mp}$, $H^0H^0$. These annihilation products
will hadronise and/or decay, eventually producing high energy muon neutrinos.
Since the density in the core of the Sun and the Earth is high, the
possibility of interactions of the annihilation products with the
surrounding medium must be considered.

In agreement with earlier work \cite{RS}, we find that light hadrons and muons
get
stopped well before they have time to decay both in the Sun and in the Earth.
Tau leptons lose hardly any energy at all and their interactions can to a good
approximation be neglected. Gauge bosons, Higgs bosons and top quarks also
decay
long before interacting. The gluon channels do not give rise to any
significant high energy muon neutrino fluxes, and can thus be neglected.
The branching ratio to neutrinos directly is close to zero for slow
neutralinos annihilating and these channels do thus not contribute.
However, the heavy hadrons ($B$'s, $D$'s, $\Lambda_b$'s etc) may or may
not interact and their interactions can not be neglected and they can not be
considered to be stopped completely either.

For the heavy hadrons we simulate the hadronisation as if the surrounding
medium were not present. Afterwards we estimate how many times the heavy
hadrons have interacted and how much energy they have lost in each
interaction. The neutrino is assumed to have lost the same fraction of its
energy as the heavy hadrons. These heavy hadron interactions are taken
care of both in the Sun and in the Earth.

The Higgs bosons decay to ordinary particles and the muon neutrino flux they
produce can be calculated from the flux that their decay products produce
\cite{Edsjo}. For the top quark, only the standard model decay $t \rightarrow
W^+b$ has been considered. In supersymmetry, the decay mode $t \rightarrow
H^+ b$ is open if the $H^+$ mass is light enough, and the flux from this
decay mode can be calculated from the fluxes of the $b$ quark and the decay
products of the $H^+$.



\begin{table*}[htb]
\newlength{\digitwidth} \settowidth{\digitwidth}{\rm 0}
\catcode`?=\active \def?{\kern\digitwidth}
\caption{Parameterization (according to Eq.\ (\protect\ref{eq:mupara})) of the
muon fluxes for different annihilation channels. With these
values the unit of the flux in Eq.\ (\protect\ref{eq:mupara}) is
m$^{-2}$(annihilation)$^{-1}$. An angular cut of $\theta < 5^{\circ}$ has been
applied.}
  \label{tab:mflx}
\begin{tabular*}{\textwidth}{@{}l@{\extracolsep{\fill}}rrrrrrrrr}  \hline
  \multicolumn{9}{c}{\bf Annihilation in the Sun} \\ \hline
  Channel & \multicolumn{1}{c}{$p_1$} & \multicolumn{1}{c}{$p_2$} &
  \multicolumn{1}{c}{$p_3$} & \multicolumn{1}{c}{$p_4$} &
  \multicolumn{1}{c}{$p_5$} & \multicolumn{1}{c}{$p_6$} &
  \multicolumn{1}{c}{$p_7$} & \multicolumn{1}{c}{$p_8$} \\ \hline
  $c \bar{c}$ & $2.39 \times 10^{-38}$ & $0.269$ & $0.00321?$ &
  $0.0598$ & $0.00937??$ &
$-0.0113?$ & $0.0119?$ & $0.0?$ \\ \hline
  $b \bar{b}$ & $6.03 \times 10^{-38}$ & $0.301$ & $0.00277?$ &
  $0.0687$ & $0.0185???$ &
$-0.00882$ & $0.0112?$ & $0.0?$ \\ \hline
  $t \bar{t}$ & $3.27 \times 10^{-37}$ & $1.31?$ & $0.000355$ &
  $0.196?$ & $0.0000399$ &
$-1.18???$ & $0.00484$ & $4.50$ \\ \hline
  $\tau^+ \tau^-$ & $2.10 \times 10^{-37}$ & $0.590$ & $0.00123?$ &
  $0.104?$ & $0.0240???$ &
$-0.222??$ & $0.00108$ & $0.0?$ \\ \hline
  $W^+ W^-$ & $2.78 \times 10^{-37}$ & $1.26?$ & $0.00110?$ & $0.229?$ &
  $0.000657?$ &
$-0.780??$ & $0.00343$ & $3.63$ \\ \hline
  $Z^0 Z^0$ & $4.14 \times 10^{-37}$ & $1.52?$ & $0.000948$ & $0.291?$ &
  $0.000368?$ &
$-1.14???$ & $0.00351$ & $3.76$ \\ \hline
\end{tabular*}


\begin{tabular*}{\textwidth}{@{}l@{\extracolsep{\fill}}rrrrrrrrr}  \hline
  \multicolumn{9}{c}{\bf Annihilation in the Earth} \\ \hline
  Channel & \multicolumn{1}{c}{$p_1$} & \multicolumn{1}{c}{$p_2$} &
  \multicolumn{1}{c}{$p_3$} & \multicolumn{1}{c}{$p_4$} &
  \multicolumn{1}{c}{$p_5$} & \multicolumn{1}{c}{$p_6$} &
  \multicolumn{1}{c}{$p_7$} & \multicolumn{1}{c}{$p_8$} \\ \hline
$c \bar{c}$ & $6.44 \times 10^{-29}$ & $0.226$ & $0.00283??$ &
  $0.0645$ & $0.00213$ & $-0.0199$ & $0.00152$ & $0.0$ \\ \hline
$b \bar{b}$ & $9.42 \times 10^{-29}$ & $0.187$ & $0.00531??$ &
  $0.0787$ & $0.00513$ & $0.0834$ & $0.00350$ & $0.0$ \\ \hline
$t \bar{t}$ & $2.48 \times 10^{-28}$ & $0.830$ & $0.0000865$ &
  $0.114?$ & $0.00581$ & $-0.667?$ & $0.0????$ & $0.0$\\ \hline
$\tau^+ \tau^-$ & $2.45 \times 10^{-28}$ & $0.152$ & $0.00533??$ &
  $0.113?$ & $0.00629$ & $0.194?$ & $0.00103$ & $0.0$ \\ \hline
$W^+ W^-$ & $2.38 \times 10^{-28}$ & $0.188$ & $-0.000672?$ &
  $0.130?$ & $0.00832$ & $0.164?$ & $0.00259$ & $0.0$ \\ \hline
$Z^0 Z^0$ & $2.87 \times 10^{-28}$ & $1.33?$ & $-0.000171?$ &
  $0.126?$ & $0.00875$ & $-1.01??$ & $0.00271$ & $0.0$ \\ \hline
\end{tabular*}
\end{table*}

\section{Neutrino interactions}
The produced muon neutrinos can interact on their way out of the Sun or the
Earth. The total neutrino-nucleon cross section is given approximately by
  $\sigma_{CC} = a (E_{\nu}/\mbox{GeV}) \times 10^{-39} \mbox{ cm}^2$
      and
  $\sigma_{NC} = b (E_{\nu}/\mbox{GeV}) \times 10^{-39} \mbox{ cm}^2$
for charged and neutral currents respectively. We have calculated the
coefficients $a$ and $b$ using the new GRV structure
functions \cite{GRV} and a $Q^2$-cut of 0.3 GeV$^2$. The result is
$a_{\nu n} = 8.81$, $a_{\nu p} = 4.51$, $a_{\bar{\nu} n} = 2.50$,
$a_{\bar{\nu} p} = 3.99$,
$b_{\nu n} = 2.20$, $b_{\nu p} = 1.97$, $b_{\bar{\nu} n} = 1.15$ and
$b_{\bar{\nu} p} = 1.14$.

One can easily find that neutrino interactions can be neglected in the
Earth, but not in the Sun. The effective thickness of protons and neutrons,
$T_p$ and $T_n$, of the Sun is calculated by using the solar
model of Bahcall et al.\ \cite{Bah} with the result
  $T_p = 1.1 \times 10^{12} \mbox{ g/cm}^2$ and
  $T_n = 3.6 \times 10^{11} \mbox{ g/cm}^2$\@.

We have approximated the Sun with a piece of homogeneous material with
the thicknesses of protons and neutrons given above. For
each produced neutrino we have calculated the mean free path for interactions
and
if an interaction has taken place, it is
chosen to be a neutral or a charged current. In case of a charged current, the
neutrino is treated as absorbed and in case of a neutral current, the
scattering is simulated with {\sc Pythia} 5.6 and the procedure is continued
with the new energy of the neutrino until the neutrino has reached the surface
of the Sun.

This differs from the approach of Ritz and Seckel \cite{RS}, which is the
common
approach, where the energy loss is considered to be continuous and
neutral currents are assumed to be much weaker than charged currents.
Neither of these assumptions is very good. In general, a neutrino only
participates in a few interactions ($\sim$0--2) on the way out of the Sun.
Hence the process is highly discrete. In Fig.\ \ref{fig:enucomp}a, the
neutrino spectrum at the surface of the Sun is shown for injection of 500 GeV
muon neutrinos and anti-neutrinos in the core of the Sun. The difference
between
the approach of Ritz and Seckel and the new method is substantial. The two
peaks in the Ritz and Seckel spectrum correspond to $\nu_{\mu}$ and
$\bar{\nu}_{\mu}$
respectively. In Fig.\ \ref{fig:enucomp}b, the neutrino spectrum weighted by
$E_{\nu}^2$ is shown for a typical annihilation channel. The two
peaks for $\nu_{\mu}$ and $\bar{\nu}_{\mu}$ can still be seen. $E_{\nu}^2
dN/dE_{\nu}$ is proportional to the differential rate of events since the
cross section for muon production (charged current) is proportional to the
energy and the range of the muon is proportional to the energy. The
difference between the two methods can clearly be seen, even though it has
been somewhat washed out when applying the methods to a neutrino
spectrum in the core of the Sun. However, the difference in the high energy
tail of the spectrum is significant and it gets important due to the
$E_{\nu}^2$-enhancement. If one uses the Ritz and Seckel approach, the error
in the total rate is significant, and of the order of 5--20\% too low with the
higher error at higher neutralino masses ($\sim 1500$ GeV).


\section{Simulation results}

The muon flux at the surface of the Earth has been calculated for different
annihilation channels and for different neutralino masses. The annihilation
channels considered are $c\bar{c}$, $b\bar{b}$, $t\bar{t}$, $\tau^+ \tau^-$,
$W^+ W^-$ and $Z^0 Z^0$ for the reasons stated earlier. The simulation has
been performed with neutralino masses of 50, 100, 150, 200, 250, 350, 500,
750, 1000 and 1500 GeV and a top quark mass of $m_{t}=150$
GeV\@. For easy use, the calculated fluxes have been parameterized
as
\begin{eqnarray}
  \lefteqn{\frac{d \Phi_{\mu}}{dz} (m_{\neu},z) = \frac{p_1 m_{\neu}^2}
  {1+\exp \left( \frac{z-p_2 \exp (-p_3 m_{\neu})-p_6} {p_4}
  \right) }  \times}  \nonumber \\
   & & \left[ 1 - \exp \left( -p_5 \frac{m_{\neu}} {z^{p_8}} \right)
\right] \exp \left( -p_7 m_{\neu} z \right)
  \label{eq:mupara}
\end{eqnarray}
where $m_{\neu}$ is the neutralino mass in GeV, $z=E_{\mu}/m_{\neu}$
and $p_1,\ldots,p_8$ are the parameters fitted to the simulation
results. The values of the fitted parameters are given in
Table~\ref{tab:mflx}\@. If one calculates the total
flux above a certain threshold, these parameterizations are accurate to $\sim
15$\% as long as the threshold is not too close to the neutralino mass
($E_{\mu}^{th} \lsim 0.2 m_{\neu}$). The parameterizations are poorest
for the $c\bar{c}$-channel and the $b\bar{b}$-channel in the Sun at high
neutralino masses ($\gsim 500$ GeV), but these channels are not expected to
dominate the flux at these high masses. The accuracy of the $\tau^+ \tau^-$,
$W^+ W^-$ and $Z^0 Z^0$ parameterizations are usually of the order of 5\% and
never worse than 10\% and these channels usually dominate the flux.

The event rate at a detector is given by
 $\Gamma_{events} = \Gamma_A A_{eff} \sum_i B_i \Phi_i$
where $\Gamma_A$ is the annihilation rate, $A_{eff}$ is the effective area of
the detector in the direction of the Sun/Earth, $B_i$ is the branching
ratio for annihilation channel $i$ and $\Phi_i$ is the muon flux (over
threshold for the detector) per unit area and annihilation. $\Gamma_A$ and
$B_i$ depend on the supersymmetric parameters chosen and are considered in
other papers (e.g.\ \cite{Kamiokande}).


\begin{figure}
  \vspace{3.7cm}
  \includegraphics{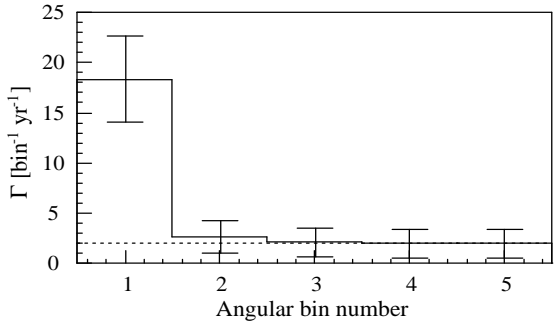}

  \caption{The expected event rate, $\Gamma$, in concentric angular bins (the
first
  one corresponding to $\pm5^{\circ}$ around the Sun and the next
  ones being of the
  same size in solid angle) for the supersymmetric model described in the
  text. The solid line corresponds to the total event rate and the
  dotted line is the background \protect\cite{GS84}. The errors
  shown are just the square root of the number of events in each bin.
  A cut of $E_{\mu}^{det}>10$ GeV is also applied.}
  \label{fig:example}
\end{figure}

\section{Example}

For a specific supersymmetric model
and a specific detector, one can calculate the event rate. Consider
a neutrino detector with an effective area of 2000 m$^2$ (a typical
size for {\sc Amanda} \cite{Amanda} in the first set-up) in the
direction of the Sun. For a Minimal Supersymmetric Model with
parameters $\mu=300$ GeV, $M_2=600$ GeV and $\tan \beta = 2.0$,
giving $m_{\neu} \simeq 250$ GeV, one gets the expected event rate
shown in
Fig.\ \ref{fig:example} where the annihilation rate and the branching
ratios are given by \cite{Lars}.
Note that this specific example is not yet excluded by other
experiments.


\section{Conclusions}

The treatment of neutrino propagation through the Sun has been improved (the
approach of Ritz and Seckel \cite{RS} was shown to give an error of up to
$\sim 20\%$) and the calculated muon fluxes are given in an easy-to-use
parameterization. The large neutrino telescopes now being built should have
good possibilities to search through new parts of the supersymmetric parameter
space.


\section*{Acknowledgments}

I am grateful to L.~Bergstr{\oo}m, P.~Gondolo and G.~Ingelman for valuable
discussions and comments.


\end{document}